\documentclass[iop]{emulateapj}
\usepackage{apjfonts}
\usepackage{graphicx}
\usepackage{natbib}
\usepackage{amsmath}
\usepackage[version=3]{mhchem}
\RequirePackage{lineno}

\begin{document}
\title{Thorium Abundances in Solar Twins and Analogues: Implications for the Habitability of Extrasolar Planetary Systems}
\author{Cayman T. Unterborn\altaffilmark{1,\textasteriskcentered}, Jennifer A. Johnson\altaffilmark{2} and Wendy R. Panero\altaffilmark{1}}
\altaffiltext{1}{School of Earth Sciences, 
The Ohio State University, 125 South Oval \indent \indent Mall, Columbus, OH 43210}
\altaffiltext{2}{Department of Astronomy and the Center for Cosmology and Astro-Particle Physics,The Ohio State University, Columbus, OH 43210}
\altaffiltext{\textasteriskcentered}{contact: unterborn.1@buckeyemail.osu.edu}
\slugcomment{Submitted to ApJ on May 12, 2014 \newline Accepted for publication on April 30, 2015}

\begin{abstract}
\indent We present the first investigation of Th abundances in Solar twins and analogues to understand the possible range of this radioactive element and its effect on rocky planet interior dynamics and potential habitability. The abundances of the radioactive elements Th and U are key components of a planet's energy budget, making up 30\% to 50\% of the Earth's \citep{Kore08,Alle01,Schu80,Lyub07,GeoNu10,Huan13}. Radiogenic heat drives interior mantle convection and surface plate tectonics, which sustains a deep carbon and water cycle and thereby aides in creating Earth's habitable surface. Unlike other heat sources that are dependent on the planet's specific formation history, the radiogenic heat budget is directly related to the mantle concentration of these nuclides. As a refractory element, the stellar abundance of Th is faithfully reflected in the terrestrial planet's concentration. We find that log $\epsilon_{\rm Th}$ varies from 59\% to 251\% that of Solar, suggesting extrasolar planetary systems may possess a greater energy budget with which to support surface to interior dynamics and thus increase their likelihood to be habitable compared to our Solar System.
\end{abstract}

\section{Introduction}
\label{sec:intro}
\indent The Earth is a dynamic planet, in which the ``heat engine'' that powers the present-day dynamics that include plate tectonics, the geodynamo in the core, and the ongoing recycling and sustentation of the Earth's atmosphere. Due to melting during the planetary formation process, the Earth has differentiated into two main compositional parts: the metallic core and the silicate-dominated mantle and crust. The metallic core contains iron, nickel, and those siderophile elements that partition into the metallic phase. The bulk silicate earth (BSE) contains the balance of the Earth's composition, including most of the oxide-forming elements. This includes K, Th, and U, whose radioactive isotopes account for 20.1 TW \citep[43\%, ][]{Huan13} of the Earth's 47 TW surface heat flow \citep{Davi10}. The remaining 26.9 TW are a result of the initial differentiation into mantle and core, secular cooling of the core and mantle and the crystallization of the inner core. Geodynamo models show that the heat flow at the core mantle boundary from the core has remained roughly constant at 9 TW \citep{Labr97}. This is $\sim$20\% of the Earth's total energy budget today and less than 10\% in the Archean, including the effects of potassium.  \\
\indent Within the BSE, however, K, Th, and U are not distributed homogeneously. These lithophile elements partition into the magma generated at subduction zones, building continental crust that accounts of 40\% of the Earth's surface, a direct consequence of plate tectonics. This process leads to heat production of 7.6 TW within the continental crust, which serves as an insulating layer and thickens the surface boundary layer. It is the remaining 12.9 TW of the radiogenic heat budget that drives mantle convection. \\
\indent Of the two main planetary heat sources (radiogenic and secular cooling), the power produced by radionuclides is entirely compositionally dependent with the radiogenic heat budget inherited upon formation. Secular cooling, in contrast, is determined by the specific formation and segregation history. Our goal here is to explore the degree to which the radiogenic heat budget can vary between terrestrial exoplanets, and consequences of formation history will be left to subsequent work. \\
\indent U and Th are both refractory elements with 50\% condensation temperatures of 1604 and 1647 K, respectively, in protoplanetary disks of Solar composition \citep{Lodd03}. The measured absolute abundances of other refractory, rock-building elements in Solar twins and analogues, stars of similar metallicity, mass and surface temperature to that of the Sun, varies between stars. Furthermore, the Sun is moderately depleted in many of the highly refractory elements (T$_{\rm Cond}$ $\textgreater$ 1500 K) of Z $\textless$ 56 \citep{Mele09}. If these differences are due to intrinsic variations in primordial abundances in planetary systems, rather than to the process of planet formation, then this suggests that the stellar abundances of U and Th may also vary within the Galaxy. In both older metal-poor stars and younger dwarfs/subgiants, wide variation in the absolute abundance of Th has been found, including other heavy r-process elements, to track Galactic chemical evolution \citep{More92,McWi95,John01,Hond04,delP05a,delP05b}. It is not currently known, however, whether these variations in Th or U are present in younger, metal-rich, planetary host stars. Any variation in these abundances will translate into similar increase or decrease in both the crust and mantle of any terrestrial planets orbiting these stars thus having implications for these heat budget of these planets and potential for dynamics.
\section{Sample and Methods}
\begin{figure}
\includegraphics[width=9cm]{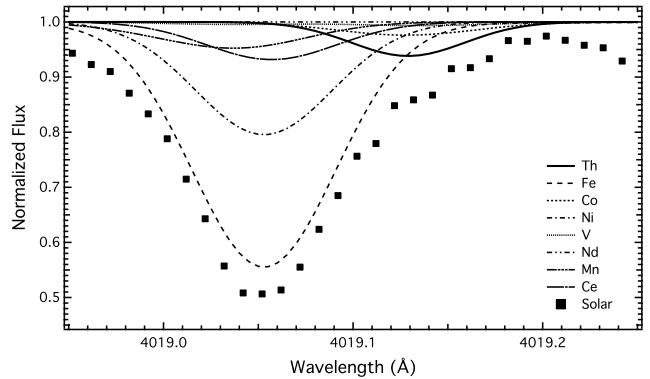}
\caption{The region between 4018.95-4019.25 \AA\ including the dominant spectral lines within this region as taken from VALD. The Solar spectrum is included for reference.}
\label{fig:Lines}
\end{figure}

\begin{deluxetable*}{lcccc|ccc|cccccc}
\tablecolumns{13}
\tablecaption{Input stellar model parameters and measured Fe, Th and Si abundances. References for stellar input model or line parameters are shown as superscripts.}
    \leavevmode
    \centering
      \epsfxsize=1cm
\tablehead{\colhead{}&\colhead{}&\colhead{}&\colhead{}&\colhead{}\vline&\colhead{\textbf{Fe$^4$}}&\colhead{\textbf{Th$^5$}}&\colhead{\# of }\vline&\multicolumn{5}{c}{\textbf{Si$^6$}}\\
\colhead{Star}&\colhead{Age (Ga)$^{2}$}&\colhead{T$_{\rm eff}$ (K)}&\colhead{$\log(g)$ (cgs)}&\colhead{[Fe/H]}\vline&\colhead{$\log \epsilon$}&\colhead{$\log \epsilon$}&\colhead{spectra}\vline&\colhead{(1)}&\colhead{(2)}&\colhead{(3)}&\colhead{(4)}&\colhead{Average}}
\cutinhead{\textbf{with observed planets}}
Sun$^1$ &4.5& - & - & -& 7.47 & 0.09 &1&-&-&-&-&-\\
Sun (This Study) &4.5& 5777 & 4.44 & 0.00 & 7.47 & 0.09 &1&7.58 & 7.60 & 7.55 & 7.55 & $7.57\pm0.02$\\
HD98649$^2$ &2.1& 5760$\pm 25$ & 4.51$\pm0.05$ & -0.01$\pm 0.02$&7.36 &  0.31 & 1& 7.54 & 7.59 & 7.54 & 7.54 & 7.565$\pm0.03$\\
HD1461$^2$&4.8 &5756$\pm 44$ & 4.37$\pm0.05$ & 0.189$\pm0.015$& 7.60$\pm0.017$  & 0.33$\pm0.05$  & 9&7.74 & 7.79 & 7.77 & 7.74 & 7.76$\pm0.02$\\
HD16417$^2$ &7.0& 5812$\pm 34$ & 4.09$\pm0.05$ & 0.094$\pm0.004$&7.53$\pm0.03$ & 0.31$\pm0.05$ & 7& 7.64 & 7.67 & 7.64 & 7.64 & 7.65$\pm0.01$\\
HD102117$^2$&5.6 & 5690$\pm22$ & 4.30$\pm0.04$ & 0.304$\pm0.003$&7.76$\pm0.02$ & 0.49$\pm0.04$ & 8& 7.85 & 7.91 & 7.95 & 7.90 & 7.90$\pm0.04$\\
HD141937$^2$&1.3 & 5900$\pm19$ & 4.45$\pm0.04$ & 0.125$\pm0.003$&7.53$\pm0.02$ & 0.44$\pm0.02$ & 3& 7.63 & 7.68 & 7.73 & 7.68 & 7.68$\pm0.04$\\
HD160691$^2$&4.6 & 5809$\pm22$ & 4.28$\pm0.04$ & 0.298$\pm0.003$&7.76$\pm0.02$ & 0.44$\pm0.09$ & 9& 7.80 & 7.85 & 7.95 & 7.85 & 7.86$\pm0.05$\\
HD10700$^3$ &-& 5312$\pm137$ & 4.59$\pm0.13$ & -0.43$\pm0.15$&6.91$\pm0.05$  & -0.14$\pm0.05$ & 4& 7.37 & 7.37 & 7.22 & 7.22 & 7.30$\pm0.08$\\
\cutinhead{\textbf{without observed planets}}
HD115169$^2$ &1.7& 5815$\pm22$ & 4.52$\pm0.05$ & -0.01$\pm0.02$&7.41 & 0.36 & 1& 7.59 & 7.59 & 7.54 & 7.59 & 7.57$\pm0.03$\\
HD146233$^2$&3.1 & 5822$\pm9$ & 4.45$\pm0.02$ & 0.051$\pm0.002$&7.48$\pm0.01$  & 0.42$\pm0.04$  & 4& 7.45 & 7.60 & 7.60 & 7.60 & 7.56$\pm0.07$\\ 
HD45346$^2$&3.8 &5745$\pm25$ & 4.47$\pm0.05$ & -0.01$\pm0.02$&7.46 & 0.16 & 1& 7.54 & 7.59 & 7.54 & 7.54 & 7.55$\pm0.02$\\ 
HD59711$^2$ &2.7& 5740$\pm15$ & 4.50$\pm0.05$ & -0.092$\pm0.02$&7.35$\pm0.05$ & 0.20$\pm0.02$ & 3& 7.56 & 7.46 & 7.46 & 7.51 & 7.50$\pm0.04$\\
HD71334$^2$&4.7 &5724$\pm15$ & 4.46$\pm0.03$ & -0.044$\pm0.02$&7.34$\pm0.04$ & 0.24$\pm0.03$ & 5 & 7.51 & 7.56 & 7.56 & 7.58 & 7.55$\pm0.03$\\
HD78534$^2$ &4.6& 5800$\pm25$ & 4.41$\pm0.05$ & 0.07$\pm0.02$& 7.49 & 0.19  & 1& 7.57 & 7.62 & 7.62 & 7.62 & 7.61$\pm0.02$\\
HD78660$^2$ &1.5& 5782$\pm29$ & 4.52$\pm0.04$ & 0.033$\pm0.02$&7.45$\pm0.01$ & 0.26$\pm0.02$ & 5 & 7.63 & 7.56 & 7.58 & 7.58 & 7.61$\pm0.03$\\
\enddata
\tablerefs{\scriptsize 1. \citet{Lodd03} 2. \citet{Baum10} 3. \citet{Mald12} 4. \citet{Snee96} 5. \citet{VALD} 6. \citet{daSi12}}
\tablecomments{\textbf{Fe:} $\lambda$ = 4019.053 \AA\ , $\log gf$ = -2.91, $\chi$ = 2.609 eV. \textbf{Th:} $\lambda$ = 4019.129 \AA\ , $\log gf$ = -0.27, $\chi$ = 0.0 eV. \textbf{Si (1):} $\lambda$ = 5517.533 \AA\ , $\log gf$ = -2.42, $\chi$ = 5.08 eV. \textbf{Si (2):} $\lambda$ = 5621.607 \AA\ , $\log gf$ = -2.61, $\chi$ = 5.08 eV. \textbf{Si (3):} $\lambda$ = 5665.563 \AA\ , $\log gf$ = -1.97, $\chi$ = 4.92 eV. \textbf{Si (4):} $\lambda$ = 5684.484 \AA\ , $\log gf$ = -1.65, $\chi$ = 4.854 eV.}
\label{tab:measured} 
\end{deluxetable*} 
\begin{figure}
\includegraphics[width=9cm]{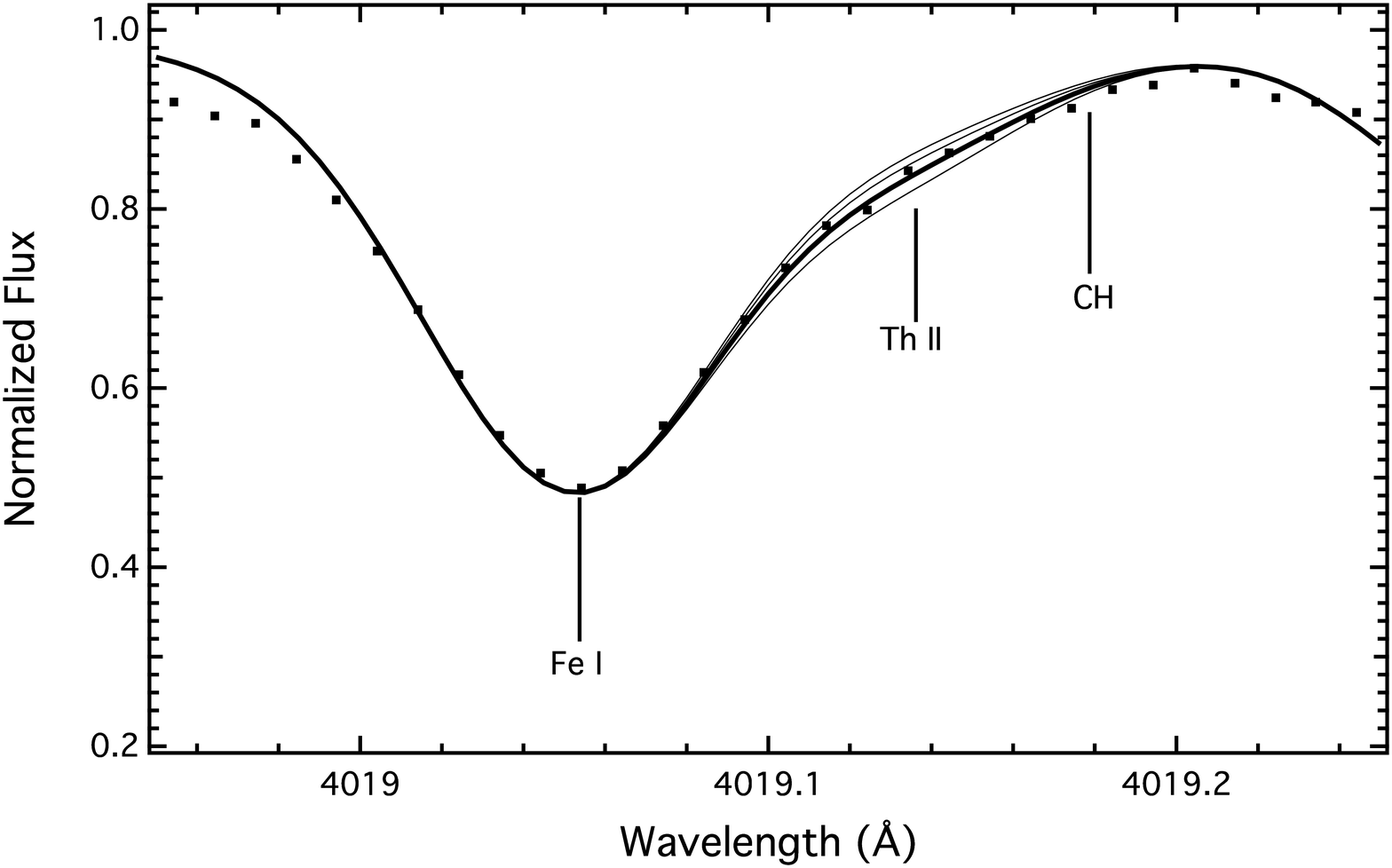}
\caption{The spectrum of HD146233 around the Th I line at 4019.129 \AA\ (dots). The synthetic spectra (solid lines) were computed adopting the stellar parameters listed in the text. Synthetic spectra were calculated using $\log \epsilon_{\rm Th}$ = 0.17, 0.25, 0.35 and 0.45. The best fit for this spectrum ($\log \epsilon_{\rm Th}$ = 0.35) is shown in bold. }
\label{fig:spectra}
\end{figure}
Our sample of fourteen stars (seven with and seven without observed planets) was selected from the publicly available HARPS archive \citep{Mayo03}. Observations were made using the 3.2 m La Silla telescope (R = 115,000). In order to minimize systematic uncertainty from input synthetic stellar models, only known Solar twins and analogues from \citet{Baum10,Mald12,Rami09} were selected (Table \ref{tab:measured}). One-dimensional, hydrostatic, plane-parallel stellar model atmospheres were obtained from the MARCS catalogue \citep{Gust08}, with raw spectra normalized using the spectrum analysis software SPECTRE \citep{Fitz87} with input stellar model parameters taken from \citet{Baum10} and \citet{Mald12}. Th abundances were measured with the line analysis software MOOG \citep{Snee73} using the blended Fe-Th line at 4019.053-4019.129 \AA\ adopting the line list of \citet{Snee96} supplemented with lines from the VALD catalogue \citep[][Table \ref{tab:measured}]{VALD}. We first fit the Fe line at 4019.053 \AA, followed by varying Th until a best fit was found (Figure \ref{fig:Abund}) as determined by a minimized $\chi ^2$ test. Our Fe abundances (Table \ref{tab:measured}, Figure \ref{fig:Abund}) are in good agreement with previous measurements for the same Solar twins and analogues, suggesting that our measurements are accurate with respect to Fe. In order to determine if any error is introduced by uncertainties in the stellar models, each abundance was remeasured adopting values 100 K, 0.05 (cgs) and 0.05 dex above and below the published value from the literature for T$_{\rm eff}$, $log(g)$ and [Fe/H] respectively (Table \ref{tab:measured}). These values are at or above the published errors for these measurements.\\
\indent There are several possible blends in our line list in addition to the well-separated Fe I line within this spectral region including contributions from Ti, V, Cr, Mn, Co, Ni, Ce, Nd, and Sm (Figure \ref{fig:Lines}). As noted in Section 3, this blending is of particular importance as it may account for differences between our measured Th abundances and those from earlier work on similar stars, particularly that of \citet{delP05a}. Of these elements, we find that only varying Co made any significant difference in our measured Th abundance, while changing the abundance of any these other element affecting only the fit to the Fe line or required variations well beyond the measured abundances of these elements \citep{delP05a} before the Th abundance was affected. Furthermore, two stars as well as the Sun overlap between our sample and that of \citet{delP05a}, with our line lists being identical. Adopting their reported Co, Fe and Th values for these stars, we are either unable to fit the measured spectra or no noticeable difference was found between their synthetic spectra and ours (Figure \ref{fig:Comp}). Furthermore, \citet{delP05a} fits $\sim$~8 Co lines for these stars, making their measured Co abundances independent of the Th value. We therefore consider any contamination to be minimal and simply adopt the scaled-solar MOOG abundances for these elements when determining Th abundance. \\
\indent Not considered by \citet{delP05a}, however, is a CH line blended at 4019.170 \AA. We find that variation in the $^{12}$C/$^{13}$C ratio was found to have a negligible effect on the spectral model for this line, and we therefore set this ratio to the Solar value of 89.4 for all measurements \citep{Aspl09}. [C/Fe] for each star was assumed to scale from Solar with metallicity as calculated by MOOG and provided a good fit to each of the $^{12}$C lines observed. \citet{Mele09} found that [C/Fe] does not scale with Solar metallicity for Solar twins and analogues, however, we find no difference in depth the $^{13}$C line when the input, MOOG C abundance is scaled by their measured variation of between 0 and 0.06 dex from Solar.\\
\indent As we discuss in Section 3, Si is the relevant rock-building element in terrestrial mantles. To measure Si abundance then, we adopt the line parameters for 4 Si lines listed in \citet{daSi12}. The values for Si abundance derived from these individual lines for our sample of 14 stars are listed in Table \ref{tab:measured}. Furthermore, we find only minor variations in derived Si abundance between these individual lines for each star in our sample. Again, each abundance was remeasured using models of 100 K, 0.05 (cgs) and 0.05 dex above and below the published value from the literature for T$_{\rm eff}$, $log(g)$ and [Fe/H] respectively, and no noticeable difference in Si abundance was found. 
\section{Results}
\begin{figure}
\centering
\includegraphics[width=7.5cm]{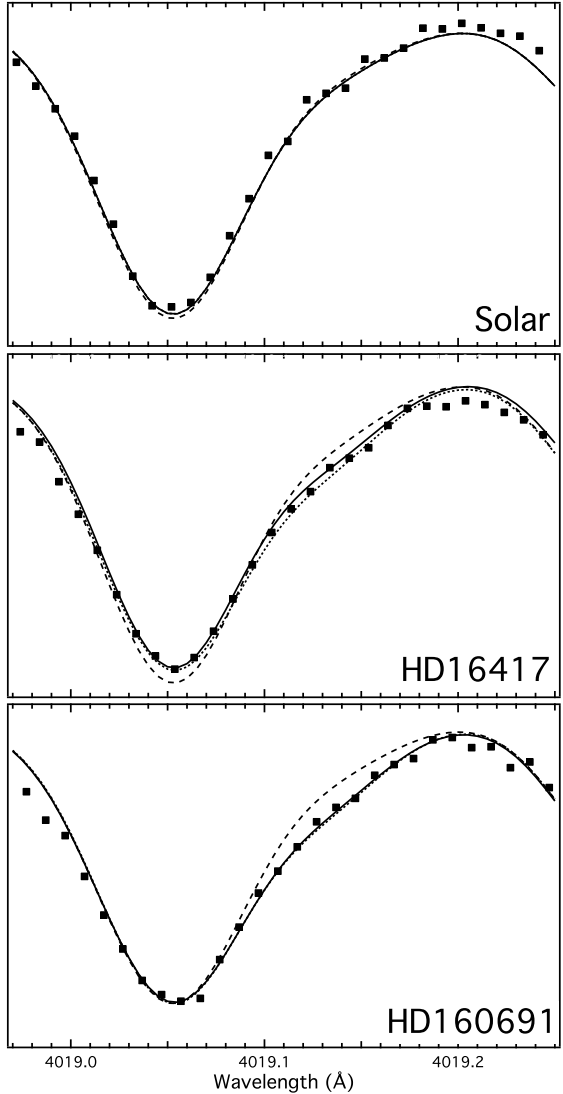}
\caption{Synthetic spectra for the three stars measured both by this study and \citet{delP05a}. Solid lines represent our synthetic spectra (Table \ref{tab:measured}), dashed lines are syntheses adopting their reported Fe, Co and Th abundances and dotted is our reported Fe and Th abundances but adopting their Co abundance. Squares are the corresponding spectra.}
\label{fig:Comp}
\end{figure}

\begin{figure*}
  \begin{center}
    \leavevmode
      \epsfxsize=15cm\plotone{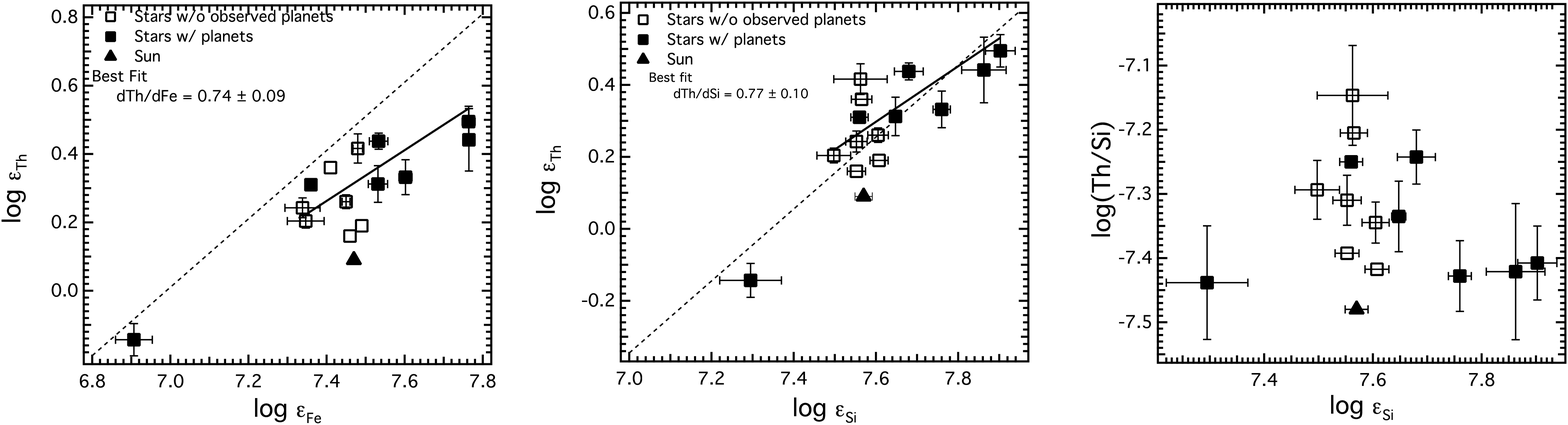}
\caption{Stars with observed planets are shown as closed squares and without as unfilled squares. The Sun is shown for reference as a triangle. \textbf{Left:} Abundance of Th as a function of Fe. The solid line represents the best fit to those stars belonging to the high Fe, high Th group and the dashed represents a one-to-one relationship. \textbf{Center:} Abundance of Th as a function of average Si abundance adopting the same legend as in Th vs Fe. \textbf{Right:} Th/Si as a function of average Si abundance. For each plot stars with observed planets are shown as closed squares and without as unfilled squares. The Sun is shown for reference as a triangle.}
\label{fig:Abund}
\end{center}
\end{figure*}

\begin{figure}
\includegraphics[width=9cm]{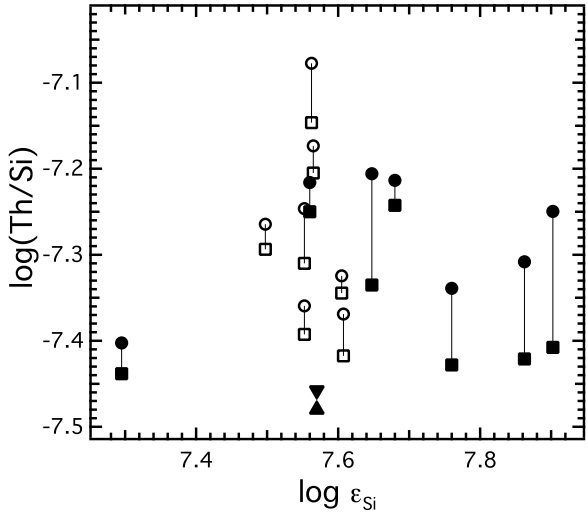}
\caption{Same as Figure \ref{fig:Abund}C with Th abundances corrected to t = 0 Ga (circles). Stars without planets are shown as open symbols, stars without as closed and the Sun as triangles. }
\label{fig:Age}
\end{figure}

\indent For this sample, we find $\log \epsilon_{\rm Th}$ between -0.14$\pm0.04$ and 0.49$\pm0.05$\footnote[1]{$\log \epsilon_{X} =  \log _{10}\left(\frac{n_X}{n_H}\right)+12$.}. Our measured Solar $\log \epsilon_{\rm Th}$ of 0.09, in agreement with the accepted literature value \citep{Lodd03}. We find that the Sun is depleted in Th/H relative to 13 stars in our sample, with one star more depleted than the Solar abundance (Figure \ref{fig:Abund}). This corresponds to a variation between 59 (+7/-7) to 251 (+24/-22)\% compared to the Solar abundance. Our total measured variation is roughly the same as \citet{delP05a}, however, in general our measured Th abundances are larger. While this enrichment is unlikely caused by our not accounting for potential contamination from other elements (Figure \ref{fig:Comp}), it may be due to our two datasets varying in both resolution (R = 48,000 vs. R = 115,000) and differences in continuum normalization. Our measured Solar Th and Fe abundances, however, are in agreement with the meteoritic value (Table \ref{tab:measured}), with \citet{delP05a} measuring depleted Th and enriched Fe abundances relative to \citet{Lodd03}. \\
\indent Differences in stellar age within our sample are another possible source of this variation. We find, however, only minor differences in Th abundance when these differences are taken into account (Figure \ref{fig:Age}). Furthermore, the half life of $^{232}$Th is 14 Ga or roughly the age of the Universe, no significant variation on account of varying ages within our sample is expected. Assuming a star's Th abundance reflects a primordial, homogeneous abundance present at the birth of every star in our Galaxy, our measured 251\% variation cannot be explained. The variation in Th abundance we observe here is likely due then to individual stars being seeded with differing amounts of r-processed material with these abundances being a lower limit on the initial variation.\\
\indent A higher Th/H in a star's atmosphere does not directly indicate a higher Th concentration in a terrestrial planet's mantle. Within the Earth, the majority of Th is present in the silicate mantle \citep{McD03}. Si is also a refractory, lithophile element in systems of Solar and near-Solar compositions \citep{Lodd03}. If we adopt then stellar Si abundance as a proxy for potential terrestrial planet mantle mass, normalizing to Th/Si allows us to compare Th concentration across planets of different silicate mantle thickness. A planet forming from materials with a high Si abundance will, to first order, result in planets with a silicate mantle accounting for a greater fraction of its radius. With this normalization then, absolute Th abundance is a function of planetary mantle mass only. The ability for a planet to convect is strongly dependent on the thickness of the mantle layer. Small planets forming with high Th/Si then may be more apt to convect due to this increase in internal heat rather than transport this heat via conduction. The Earth's BSE has an effective log(Th/Si)$\sim$-7.34 \citep{McD03}, a 28\% enrichment of Th relative to the Sun \citep[log(Th/Si)=-7.45, ][]{Lodd03}. This may be a consequence of small amounts of Si incorporated into the Earth's core, constrained via joint mineralogical and seismic models to be no more than 4.5 wt\% (best fit 1.9 wt\%). Accounting for Si in the core, the Earth's log(Th/Si) is greater than or equal to -7.38 (a 10\% decrease). This decrease is small compared to our observed 251 (+24/-22)\% variation in Th abundances and thus normalizing to Si provides us with a good proxy for planetary mantle Th concentration.\\
\indent The abundance of Si for majority of our sample falls between $\log \epsilon_{Si}$ of 7.5 and 7.9 including that of the Sun (Figure \ref{fig:Abund}). Within this range, however, Th/Si varies by a factor of 2.6, suggesting that the observed variation in Th indeed correlates with an increased concentration of Th within the silicate species. We further find a positive correlation of Th with Si ($d$Th/$d$Si = 0.77). We do not observe any variation of Th/Si with T$_{\rm eff}$ and therefore do not suspect any significant selection effects in our sample. We see this enrichment relative to Solar in stars both with and without observed planets, suggesting that although the planet-hosting stars contain known gas giants, these enrichments are not indicative of the infall of refractory material back onto the parent star. The variance in our observations suggest that any terrestrial planets present in these systems may be more or less dynamic due to variations in their energy budget from radiogenic heat as compared to the Solar System.
\subsection{Thermal Effects of Th budget on Planetary Dynamics}
\begin{figure}
\includegraphics[width=8.5cm]{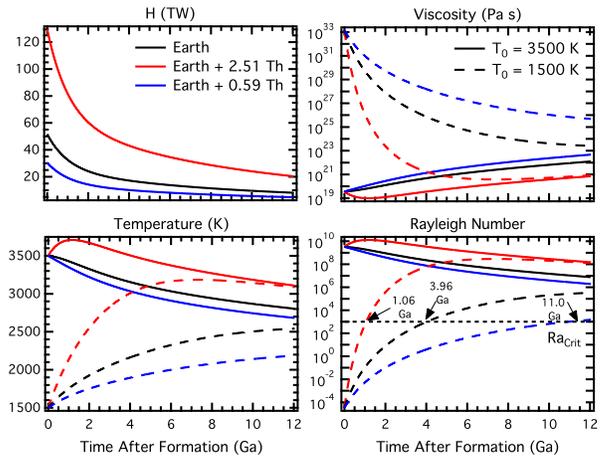}
\caption{Power produced from radiogenic sources, viscosity, temperature and Rayleigh number for a 1 M$_{\oplus}$, 1 R$_{\oplus}$ planet of Earth-like composition with 198.3 (red), 79 (black) and 46.6 (blue) ppb Th up to t = 12 Ga. Models were run with an initial temperature of 1500 K (dashed) and 3500 K (solid) with $\beta$ = 0.2. The approximate critical Rayleigh number (10$^3$) is shown for reference as a dotted horizontal line.\newline}
\label{fig:HighH}
\end{figure}

\indent The nucleosynthetic origins of Th and U are observationally and theoretically correlated \citep{Gori01,Freb07} as well as directly measured in meteorites and crustal minerals in the Earth \citep{Lodd03,Urey52, RudG03} and both must be considered in the total heat budget of a planetary mantle. If we adopt the average measured present-day BSE value of Th/U = 4 and scale the current BSE Th abundance from geoneutrino studies of 79 ppb \citep{Huan13} by the maximum Th/Th$_{\rm Sun}$ in our data set, we find a planet's energy radioactive heat budget will be on average $\sim$ 26 TW greater than the Earth over 12 Ga after planetary formation (Figure \ref{fig:HighH}).
\subsection{Thermal Model}
We adopt a single layer parameterized convection model \citep{McNa00} to determine the effects on a planet's dynamic state due to this change in a planet's internal energy budget. This model relies on the system's Rayleigh number as a measure of convective vigor. While a parameterized convection model is not as accurate as a full, 3D convection model, it allows us to explore the general dynamical effects of incorporating a varying proportion of heat-producing radionuclides into a planetary mantle and to guide further investigation using more exact models. There is no mechanism to induce or sustain plate tectonics without interior convection and although plate tectonic-like regimes are a more complex question of fault strength, surface gravity, and the presence of liquid water \citep{Vale09, Vale07,Crow11,VanH11, Tack13}, the Rayleigh number provides, to first order, a measure of whether interior mantle convection, and thus the possibility of tectonics will occur in these planets and be potentially habitable. \\
\indent The Rayleigh number ($Ra$) of a system as a function of time, $t$, is:
\begin{equation}
\label{Rayleigh}
Ra(t)=\frac{g\rho_{0}^2\alpha \Delta T(t)D^{3}C_{p}}{k\eta _{0}(T)}
\end{equation}
where $g$ is the acceleration due to gravity, $\rho_{0}$ is the average density of the mantle, $\alpha$ is the thermal expansivity of the materials present, $\Delta T(t)$ is the change in temperature across the convecting layer, $D$ is the thickness of the convecting layer, $C_{p}$ is the specific heat of the system, $k$ is the thermal conductivity and $\eta_{0}$ is an average viscosity. Above the critical Rayleigh number of 10$^3$ \citep{Schu79,Schu80}, mantle interior heat will be more efficiently transported via convection, with conduction dominating below the critical value.\\ 
\indent A parameterized convection model relates the efficiency of heat extraction from the planet to the vigor of convection, which, in turn, is dependent upon the internal temperature. The temperature change in a planetary body is quantified through conservation of energy, such that:
\begin{equation}
\label{COE}
M_mC_{p}\frac{dT}{dt} = H(t) - Q(t)
\end{equation}
where $H(t)$ is the amount of heat generated within the mantle, and $Q(t)$ is the heat extracted from the outer surface and $M_m$ is the mass of the mantle. Heat extraction at the surface is limited by the thickness of the surface boundary layer, $\delta$. We adopt the relationship:
\begin{equation} 
\label{delta}
\delta = D\left(\frac{Ra_{crit}}{Ra(t)}\right)^\beta
\end{equation}
where $\beta$ is a scaling factor that relates the vigor of convection to the surface heat flux, $Q(t)$. The value of $\beta$ for the Earth is a matter of debate, but is general considered to be between 0 and 0.33 and we adopt a median value of 0.2 for this model. Equation (\ref{delta}) may be rewritten in terms of the surface heat flow in Watts as $Q(t)=4\pi r^2*k\Delta T/\delta$:
\begin{equation}
\label{heat}
Q(t)=4\pi r^2\frac{k\Delta T(t)}{D}\left(\frac{Ra(t)}{Ra_{crit}}\right)^{\beta} 
\end{equation}
where $r$ is the radius of the planet.\\
\indent Combining equations (\ref{Rayleigh} - \ref{heat}) yields:
\begin{equation}
\label{DiffEQ}
M_mC_{p}\frac{dT}{dt} = H(t)- 4\pi r^2\frac{k\Delta T(t)}{D}\left(\frac{g\rho_{0}^2\alpha \Delta T(t)D^{3}C_{p}}{k\eta _{0}(T)Ra_{crit}}\right)^\beta
\end{equation} \\
\noindent This captures the basic physics of the feedback loop between temperature, internal viscosity, and surface heat loss, by which the increased vigor of convection increases surface heat loss, thereby reducing internal temperatures, increasing the viscosity and thus limiting internal convection \citep{Chri85}. This model assumes that any heat transported to the surface is removed from the system with no insulating thermal boundary layer at the top of the convection zone (i.e. continents). While not strictly valid in the presence of an insulating atmosphere, this assumption allows us to see the first-order effects of an increase in overall thermal conductivity and viscosity profile of a planet with increased Th and U abundance. It also does not take into account any feedback from tectonic recycling of cooler surface plates into the hot mantle, and this model is therefore meant as a first order approximation of the thermal history of a planet to determine the effects of composition on bulk thermal transport.  \\
\indent We adopt Earth-like values for $C_{p}$, $k$ and $\alpha$ (Table \ref{tab:Model}). In the high temperature limit, $C_{p}$ is a constant, and $\alpha$ and $k$ will vary by less than an order of magnitude due to bulk mineralogical differences and temperature effects, thus expressing very little variability in equation \ref{DiffEQ}. In calculating $\Delta T(t)$, we adopt $\Delta T(t) = T(t) - T_{s}$, where $T_{s}$ is the temperature at the top of the convecting layer. In our model we hold $T_{s}$, at a constant temperature of 273 K.  \\
\indent The thermal evolution model is dependent upon an initial average temperature of the planet's interior ($T_{0}$). This reflects the formation history of the planet and is primarily dependent upon two sources of heat: gravitational energy converted to heat via both planetary accretion and the segregation of a central core. Therefore we assert that varying $T_{0}$ serves as a proxy for the amount of latent primordial heat present via these history-dependent mechanisms, with high $T_{0}$ corresponding to either planets that formed quickly (and thus retained much of their accretionary heat) or segregated a large core relative to their overall volume.
\subsubsection{Mantle Heat Generation}
\indent Heat production, $H(t)$, arises from the sum of the energy produced via the radioactive decay of $^{235}$U, $^{238}$U, $^{232}$Th and $^{40}$K. We adopt Earth-like mantle heat production as our basis for scaling Th abundance with present-day values of Th/U = 4 and K/U = 10$^{4}$ and a BSE value of $^{232}$Th = 79 ppb of \citet{Huan13} with $^{238}$U/U = 0.9928, $^{40}$K/K = 1.19*10$^{-4}$ \citep{Turc80}. These values, as well as the respective ratios, are scaled back in time to determine an $H(t=0)$ of 51.6 TW consistent with the back extrapolation in time from the present value of $H(4.5 Ga)$ = 16.4 TW. While a significant fraction of Th is concentrated in Earth's continental crust, this fractionation is a direct consequence of melting associated with plate tectonics and mantle convection modeled here. Therefore, we choose to model initially the influence of variation in this primordial Th abundance on planetary mantle dynamics rather than the more complex question of sequestering some of these radionuclides in the crust due to plate tectonics.
\subsubsection{Viscosity}
\indent Power-law creep is the dominant creep mechanism in Earth's mantle, which is dependent upon the convective driving stress.  For simplicity sake, we consider a diffusion creep model \citep[e.g. ][]{Kara93} to capture the broad temperature and material dependence on viscosity.  We assume that perovskite viscosity scales relative to dry olivine diffusion creep using the equation for strain rate, $\dot{\varepsilon}=A(\sigma/\mu)(b/d)^{m}\exp[-(E^{*}+PV^{*}/RT)]$, where $\mathit{A}$ is the preexoponential factor, $\mathit{\sigma}$ is the shear stress, $\mathit{\mu}$ is the shear modulus, $\mathit{b}$ is the length of the Burgers vector, $\mathit{d}$ is the grain size, $\mathit{m}$ is the grain size exponent, $\mathit{E^{*}}$ is the activation energy, $\mathit{V^{*}}$ is the activation volume, $\mathit{R}$ is the gas constant, $\mathit{P}$ is the pressure and $\mathit{T}$ is the temperature. The effective viscosity is then:
\begin{equation}
\label{Visc}
\eta(T) = \eta_{0}exp\left[\left(E^{*}+PV^{*}\right)/RT\right]
\end{equation}
where $\eta_{0}$ is the viscosity coefficient  for a fixed driving stress. Viscosity model parameters are as in Table \ref{tab:Model}. All viscosity calculations were calculated at a constant pressure of 86 GPa, which roughly corresponds to the pressure at one half of the Earth mantle's radius.
\begin{deluxetable}{lcc}
\tablecolumns{3}
\tablecaption{Rayleigh number parameters adopted in our model}
\tablehead{\colhead{Parameter}&\colhead{Value}&\colhead{Reference}}
\startdata
r (km) & 6371 & -  \\
D (km) &2873&1 \\
g (m s$^{-2}$) & 9.81&1\\
$\rho_{0}$ (kg m$^{-3}$)&4400&1\\
$\rm M_m$ (kg) & 4.04$*10^{24}$&1\\
$\rm C_p$ (J K$^{-1}$ kg$^{-1}$)& 1250 & 1 \\ 
$\alpha$ (K$^{-1}$)&3$*10^{-5}$&1\\
k (W m$^{-1}$ K$^{-1}$)&5.6&1\\
A(s$^{-1}$)&2.67$*10^{17}$&scaled to diffusivities \\
m&2.5&assumed same as olivine\\
d (m)&0.005&equal to xenolith grains\\
b (m)&6$*10^{-10}$&scaled to olivine lattice\\
$\mu$(GPa)&184&2\\
$\eta_{0}$ (Pa s)&3.13$*10^{9}$&calculated\\
E$^{*}$ (kJ mol$^{-1}$) & 502.54&3\\
V$^*$ (cc mol$^{-1}$)&2.0&4\\
\enddata
\tablerefs{\scriptsize 1. \citet{McNa00} 2. \citet{Kepp08} 3. \citet{Dobs08} 4. \citet{Holz05}}
\label{tab:Model}
\end{deluxetable}
\subsection{Results of Thermal Model}
\indent Solving Equation \ref{DiffEQ} using a 4th order Runge-Kutta method, we find that for planets forming with a high initial temperature (3500 K), the Rayleigh number remains above the critical value for convection regardless of Th concentration over the entire time period considered here. For planets forming with a low initial temperature (1500 K), characteristic of a slow coalescence from planetesimals, however, a factor of 2.51 increase in Th abundance reduces the time at which the planet's Rayleigh number rises above the critical value by $\sim$3 Ga. When the Th abundance is decreased by a factor of 0.59 this crossover time increases by 7 Ga compared to an Earth-like case. These results, however, do not consider the additional potential variability due to radioactive $^{40}$K, a moderately volatile element with a 50\% condensation temperature = 1006 K \citep{Lodd03}. While an increase in a planets $^{40}$K abundance will further increase the chances of convection in a planet, stellar abundance measurements of K will not as directly inform its contribution to the potential heat budgets and habitability of terrestrial planets as compared to the refractory elements U and Th.
\section{Conclusion}
The measurements presented here are the first measurement of a planetary system's potential energy budget and potential for planetary interior dynamics. Mantle convection, as the underlying mechanism driving plate tectonics on the Earth, recycles and regenerates crustal, oceanic and atmospheric material \citep{Crow11,Kore11}. This surface-to-interior process regulates the global carbon and water cycles, which in turn, aid in creating a habitable surface \citep{Slee01}. We find that some stellar systems possess a greater Th abundance compared to Solar and due to the refractory nature of Th, this increase will be reflected in the mantles and crusts of any resulting terrestrial planets. Planetary systems with higher Th/Si ratios at t = 0 Ga than Solar suggest that these systems possess larger energy budgets. This supports interior dynamics, and increases the likelihood for carbon and water cycling between the surface crust and planetary interior, thus broadening the range of planets which may support habitable surfaces.\acknowledgments
This work is supported by NSF CAREER grant EAR-60023026 to WRP.
\bibliographystyle{apj}

\bibliography{Unterborn_Thorium}
\end{document}